\documentstyle[prl,aps,preprint]{revtex}

\begin{document}
\renewcommand{\theequation}{\arabic{section}.\arabic{equation}}

\widetext
\draft
\title{
Sine-Gordon Field Theory\\
for the Kosterlitz-Thouless Transitions on 
Fluctuating Membranes
}
\author{Jeong-Man Park and T. C. Lubensky}
\address{
Department of Physics, University of Pennsylvania, \\
Philadelphia, PA 19104
}
\maketitle

\begin{abstract}
In the preceding paper, we derived Coulomb-gas and sine-Gordon Hamiltonians
to describe the Kosterlitz-Thouless transition on a fluctuating surface.
These Hamiltonians contain couplings to Gaussian curvature not found
in a rigid flat surface. In this paper, we derive renormalization-group
recursion relations for the sine-Gordon model using field-theoretic
techniques developed to study flat space problems.
\end{abstract}

\section{The sine-Gordon Hamiltonian}
The Hamiltonian and associated partition function describing
$p$-atic order on a fluctuating surface were derived and
analyzed in the Coulomb gas model in the previous paper \cite{jp-lub}.
The renormalization-group (RG) recursion relations for $K$, $\kappa$, 
and $y$ were also derived
from the Coulomb gas model.
In addition, we described there how the two-dimensional Coulomb gas model can
be transformed into a sine-Gordon field theory.
\par
Here we describe to what extent these results can be verified by
a well-controlled renormalization procedure
based on the sine-Gordon field theory.
The two-dimensional Coulomb gas model can be converted into the
sine-Gordon Hamiltonian via a Hubbard-Stratonovich transformation.
The advantage of this transformation is that it makes available to us 
standard field theory diagrammatics and renormalization procedures
\cite{amit-book,amit}.
This opens up a systematic way of obtaining results for the RG
recursion relations.
\par
The $p$-atic membrane partition function in the sine-Gordon field
theory is written as
\begin{equation}
{\cal Z} = \int{\cal D}\phi{\cal D}{\bf R} e^{-\beta{\cal H}_{\kappa}
-\beta{\cal H}_{\phi}-i(p/2\pi)\int d^{2}u\sqrt{g} S\phi},
\label{sgham}
\end{equation}
where $\beta$ is the inverse temperature,
\begin{equation}
\beta{\cal H}_{\kappa} = \frac{1}{2}\beta\kappa\int d^{2}u\sqrt{g} H^{2},
\label{bending}
\end{equation}
and
\begin{equation}
\beta{\cal H}_{\phi} = \frac{1}{2\beta K}\left(\frac{p}{2\pi}\right)^{2}
\int d^{2}u\sqrt{g} g^{\alpha\beta}\partial_{\alpha}\phi\partial_{\beta}\phi  
         - \frac{2y}{a^{2}}\int d^{2}u\sqrt{g} \cos\phi 
\label{sgft}
\end{equation}
with $\case{1}{2}H=\case{1}{2}K^{\alpha}_{\alpha}$ the mean curvature
and $S=\det K^{\alpha}_{\beta}$ the Gaussian curvature.
In the Monge gauge, the metric tensor $g_{\alpha\beta}$ is written as
\begin{equation}
g_{\alpha\beta} = \partial_{\alpha}{\bf R}\cdot\partial_{\beta}{\bf R}
= \left(  \begin{array}{cc}
          1+(\partial_{x}h)^{2} & \partial_{x}h\partial_{y}h  \\
          \partial_{x}h\partial_{y}h  & 1+(\partial_{y}h)^{2} 
          \end{array}
  \right),
\end{equation}
and the curvature tensor $K_{\alpha\beta}$ is
\begin{equation}
K_{\alpha\beta} = {\bf N}\cdot D_{\alpha}D_{\beta}{\bf R}
= \frac{1}{\sqrt{1+(\nabla h)^{2}}} \left( \begin{array}{cc}
             \partial_{x}\partial_{x}h & \partial_{x}\partial_{y}h   \\
             \partial_{y}\partial_{x}h & \partial_{y}\partial_{y}h 
             \end{array}  \right),
\end{equation}
where $(\nabla h)^{2} = (\partial_{x}h)^{2} + (\partial_{y}h)^{2}$.
To lowest order in $h$, 
\begin{equation}
H = g^{\alpha\beta}K_{\alpha\beta} =
\partial_{x}\partial_{x}h + \partial_{y}\partial_{y}h = \nabla^{2}h,
\end{equation}
and
\begin{equation}
S = \det g^{\alpha\beta}K_{\alpha\beta} = \frac{1}{2}(
\nabla^{2}h\nabla^{2}h - \partial_{i}\partial_{j}h\partial_{i}\partial_{j}h).
\end{equation}
\par
In performing a perturbation calculation with Hamiltonian (\ref{bending})
and (\ref{sgft}) in two dimensions, one faces infrared as well as ultraviolet
divergences.
For the infrared regularization, we introduce a tension term 
\begin{equation}
\beta{\cal H}_{\sigma} = \beta\sigma\int d^{2}u\sqrt{g}
\cong \beta\sigma\int d^{2}x \left( 1+\case{1}{2}(\nabla h)^{2} \right)
\label{sur-ten}
\end{equation}
into the
bending Hamiltonian (\ref{bending}) and a mass term 
\begin{equation}
\beta{\cal H}_{m} = \frac{1}{2\beta K}\left(\frac{p}{2\pi}\right)^{2}
\int d^{2}u\sqrt{g} m^{2}\phi^{2}
\end{equation}
into the sine-Gordon
Hamiltonian (\ref{sgft}). 
Our infrared treatment is similar to the infrared regularization in Ref.
\cite{amit}. 
After perturbation theory is summed, we shall see that our 
renormalization group  recursion
relations have a well-defined limit when surface tension and mass are
set to zero.
Thus we will carry out our calculations in the presence of the tension $\sigma$
and the mass $m$, and set $\sigma\rightarrow 0$ and $m\rightarrow 0$ 
at the end.
\par
The ultraviolet regularization is introduced by a short-distance cut-off,
$a$, in the free propagators in coordinate space:
\begin{eqnarray}
G^{0}_{\phi\phi}(x,a) & = & \left. \frac{4\pi^{2}\beta K}{p^{2}}
      \int\frac{d^{2}q}{(2\pi)^{2}}\frac{e^{iq\cdot y}}
      {q^{2}+m^{2}} \right|_{y^{2}=x^{2}+a^{2}}   \nonumber  \\
   & = & \frac{4\pi^{2}\beta K}{p^{2}}
      \frac{1}{2\pi} K_{0}[m\sqrt{x^{2}+a^{2}}]  \nonumber  \\
   & \sim & -\frac{1}{4\pi}\frac{4\pi^{2}\beta K}{p^{2}}
      \ln (cm^{2}(x^{2}+a^{2})) \;\;, \;\;\;
            m\sqrt{x^{2}+a^{2}} \ll 1,
\end{eqnarray}
and
\begin{eqnarray}
G^{0}_{\nabla h\nabla h}(x,a) & = & \left. \frac{1}{\beta\kappa}
      \int\frac{d^{2}q}{(2\pi)^{2}}\frac{e^{iq\cdot y}}
      {q^{2}+\sigma} \right|_{y^{2}=x^{2}+a^{2}}   \nonumber  \\
   & = & \frac{1}{\beta\kappa}
      \frac{1}{2\pi} K_{0}[\sigma^{1/2}\sqrt{x^{2}+a^{2}}]  \nonumber  \\
   & \sim & -\frac{1}{4\pi}\frac{1}{\beta\kappa}
      \ln (c\sigma(x^{2}+a^{2})) \;\;, \;\;\;
            \sqrt{\sigma(x^{2}+a^{2})} \ll 1.
\end{eqnarray}
The Fourier transforms of these functions are
\begin{eqnarray}
G^{0}_{\phi\phi}(q,a) & = & \frac{4\pi^{2}\beta K}{p^{2}}
       \frac{a\sqrt{q^{2}+m^{2}}}{q^{2}+m^{2}}
       K_{1}[a\sqrt{q^{2}+m^{2}}]  \nonumber  \\
   & \sim & \frac{4\pi^{2}\beta K}{p^{2}}
       \frac{1}{q^{2}+m^{2}} \;\; , \;\;\; a \rightarrow 0,
\end{eqnarray}
and
\begin{eqnarray}
G^{0}_{\nabla h\nabla h}(q,a) & = & \frac{1}{\beta\kappa}
       \frac{a\sqrt{q^{2}+\sigma}}{q^{2}+\sigma}
       K_{1}[a\sqrt{q^{2}+\sigma}]  \nonumber  \\
   & \sim & \frac{1}{\beta\kappa}
       \frac{1}{q^{2}+\sigma} \;\; , \;\;\; a \rightarrow 0,
\end{eqnarray}
where $K_{0}(x)$ and $K_{1}(x)$ are the conventional Bessel functions 
\cite{Bessel}
and $c = e^{2\gamma}/4$ with $\gamma$ the Euler constant.
We use this ultraviolet regularization rather than the sharp cut-off
in momentum space since our calculations can be easily evaluated in 
coordinate space \cite{amit,zinn,drouffe}.
\setcounter{equation}{0}
\section{RG recursion relations}
In order to establish the RG recursion relations for the
bending rigidity, the hexatic rigidity, and the fugacity of the 
fluctuating hexatic membranes, we study the renormalizations of
the two-point vertex functions
$\Gamma^{(2)}_{\phi\phi}(q)$ and $\Gamma^{(2)}_{hh}(q)$ for the
total Hamiltonian in (\ref{sgham}).
We have two independent sets of expansion parameters which are small.
One consists of the single parameter, 
$(\beta\kappa)^{-1}$; the other consists of the expansion parameters of
the 2D XY model: the deviation of the hexatic rigidity 
from the fixed point value, $\delta = (\pi\beta K/2p^{2})-1$, and the fugacity
of the disclinations, $y$.
We will carry out calculations to lowest order in temperature
$[(\beta\kappa)^{-1}]$ and combined second order in $\delta$ and $y$.
Thus we divide the renormalization scheme into two separate parts;
first we obtain the vertex functions $\tilde{\Gamma}^{(2)}$ 
for $y=0$ up to first order in
temperature, then we turn on the effect of the disclinations
$(y\neq 0)$ and calculate the vertex functions $\Gamma^{(2)}$ 
up to combined second order in $\delta$ and $y$.
\par
\begin{figure}
\centerline{\psfig{figure=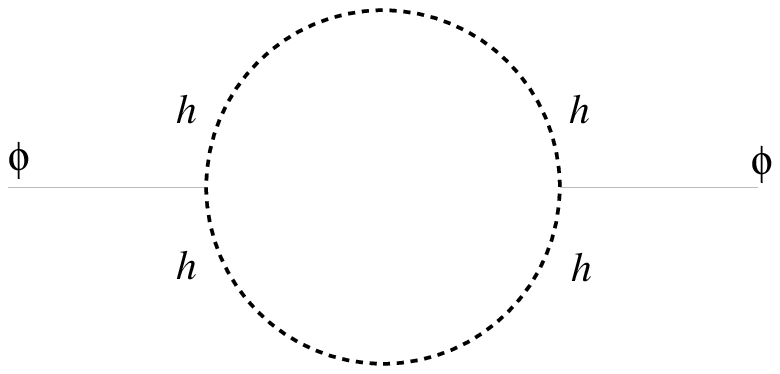}}
\caption{Diagram contributing to the shift of the stiffness.
}
\label{gpp}
\end{figure}

Starting with $y=0$, we calculate the two-point vertex function
$\tilde{\Gamma}^{(2)}_{\phi\phi}$ for the sine-Gordon field $\phi$
to lowest order in temperature. The only diagram contributing to
$\tilde{\Gamma}^{(2)}_{\phi\phi}$ in this limit is shown in Fig. \ref{gpp}.
It arises from the coupling of $\phi$ to $h$.
When $\sigma = 0$, the resulting vertex can be written as
\begin{equation}
\tilde{\Gamma}^{(2)}_{\phi\phi}(q) = \left( \frac{p^{2}}{4\pi^{2}\beta K} 
 + \frac{p^{2}}{4\pi^{2}}\frac{3}{32\pi}\frac{1}{(\beta\kappa)^{2}}
\right) q^{2} + \frac{p^{2}}{4\pi\beta K}m^{2}.
\end{equation}
Thus, the effect of height fluctuation is to shift the hexatic
rigidity from $K$ to ${\overline K}$, where to leading order in temperature
[$(\beta\kappa)^{-1}$]
\begin{equation}
(\beta{\overline K})^{-1} = (\beta K)^{-1} + 
\frac{3}{32\pi}(\beta\kappa)^{-2}.
\end{equation}
\par
\begin{figure}
\centerline{\psfig{figure=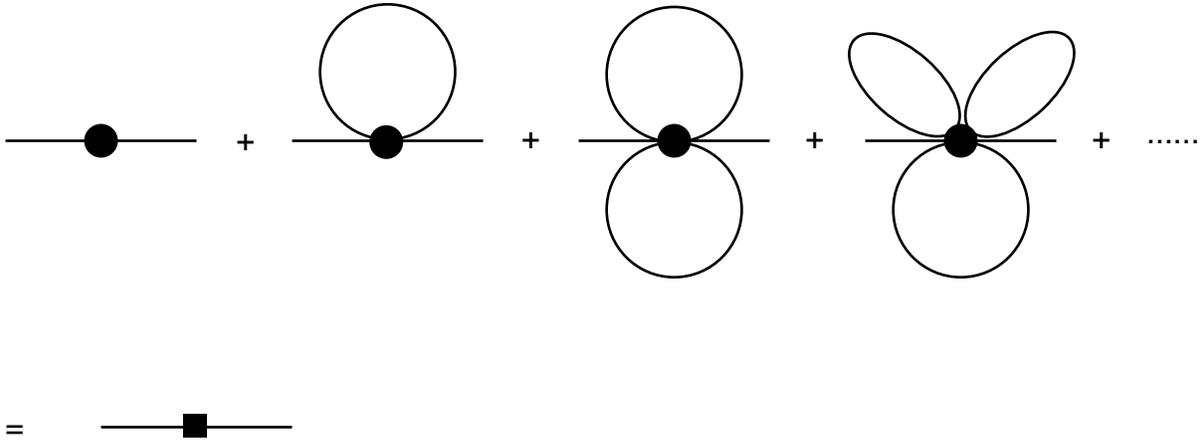}}
\caption{The circular dot represents the $\cos$ interaction and
the square dot represents the sum of
tadpole diagrams.
}
\label{fugacity}
\end{figure}

Now taking into account the effect of the disclinations, we calculate
$\Gamma^{(2)}_{\phi\phi}$ up to second order in $\delta$ and $y$.
The first step in the computation of $\Gamma^{(2)}_{\phi\phi}$ 
is the summation of tadpoles.
By expanding the $\cos\phi$ interaction of the sine-Gordon Hamiltonian
in powers of $\phi$, one generates diagrams composed of vertices
of all even orders and bare propagator $G^{0}_{\phi\phi}(q,a)$. 
In Ref. \cite{Coleman}, Coleman noted that only diagrams containing 
tadpoles are
ultraviolet-divergent logarithmically and these divergences can be removed
by renormalizing the fugacity $y$. 
Tadpoles diagrams are shown in Fig.~\ref{fugacity}.
The effect of tadpoles is to renormalize $y$. Any diagram can be described
as a diagram without tadpoles plus tadpoles adjointed.
The effect of adding an arbitrary number of tadpoles to a bare vertex
is to multiply each vertex by 
\begin{equation}
e^{-\frac{1}{2}G^{0}_{\phi\phi}(x=0,a)},
\end{equation}
thereby renormalizing $y$ :
\begin{eqnarray}
ya^{-2} & \rightarrow & ya^{-2}e^{-\frac{1}{2}G_{0}(x=0,a)}  \nonumber  \\
        & = & ycm^{2}(cm^{2}a^{2})^{\frac{\pi\beta\bar{K}_{A}}{2n^{2}} - 1}
              \equiv y{\cal A},
\end{eqnarray}
where the last equation defines ${\cal A}$.
\par
\begin{figure}
\centerline{\psfig{figure=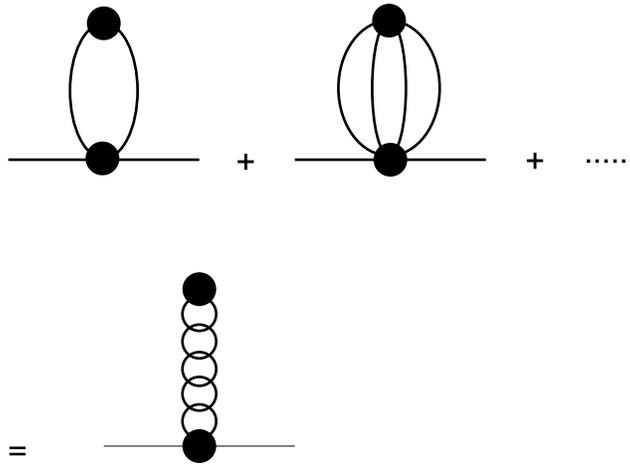}}
\caption{The sum of even numbers of intermediate propagators is
represented by the vertical line of circles.
}
\label{cosh}
\end{figure}
\begin{figure}
\centerline{\psfig{figure=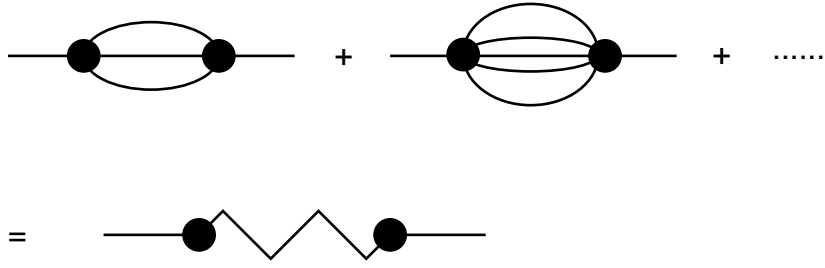}}
\caption{The sum of odd numbers of intermediate propagators is represented by
the sawtooth line.
}
\label{sinh}
\end{figure}
\begin{figure}
\centerline{\psfig{figure=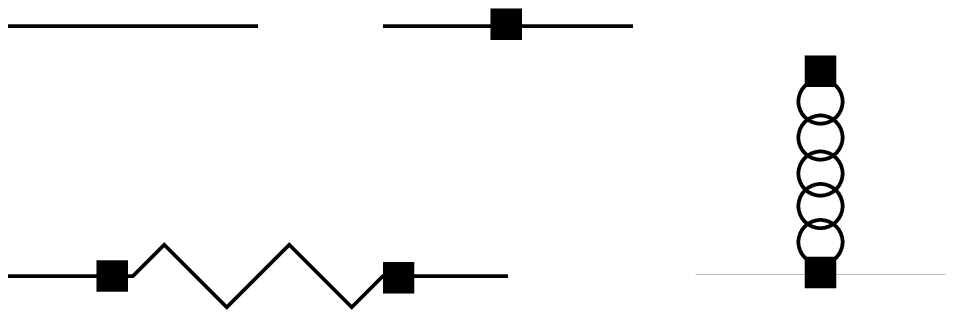}}
\caption{Diagrams contributing to $\Gamma^{(2)}_{\phi\phi}$ up to
second order in $y$. 
The square dot represents the sum of
tadpole diagrams in Fig. {\protect \ref{fugacity}}, the verticle line of circles
the sum of even numbers of propagators shown in Fig. {\protect \ref{cosh}} and
the sawtooth line the sum of odd numbers of propagators shown in Fig. 
{\protect \ref{sinh}}.
}
\label{phiphi}
\end{figure}

Next we introduce some diagrammatic notation. The sum of even
numbers of intermediate propargators is
$[\cosh G^{0}_{\phi\phi}(x) -1]$, and 
the sum of odd numbers of intermediate propargators
is $[\sinh G^{0}_{\phi\phi}(x) - G^{0}_{\phi\phi}(x)]$. 
The diagrams corresponding to 
these sums are shown in Fig. \ref{cosh} and Fig. \ref{sinh}.
Using this notation, all diagrams contributing to 
$\Gamma^{(2)}_{\phi\phi}$ up to
second order in $\delta$ and $y$ are shown in Fig. \ref{phiphi}.
Their sum can be written as 
\begin{eqnarray}
\Gamma_{\phi\phi}^{(2)}(q) & = & \frac{p^{2}}{4\pi^{2}\beta{\overline K}}
       (q^{2} + m^{2}) + 2y{\cal A}
       \nonumber  \\
 & & - \left( y{\cal A}
       \right)^{2}\int d^{2}x \left[ e^{iq\cdot x}[\sinh G^{0}_{\phi\phi}(x) 
       - G^{0}_{\phi\phi}(x)]
       - [\cosh G^{0}_{\phi\phi}(x) -1] \right].
\end{eqnarray}
To second order in $\delta$ and $y$, the first order term in $y$ contributes 
a factor of $\ln a$ :
\begin{equation}
y{\cal A} =
ycm^{2}(cm^{2}a^{2})^{\frac{\pi\beta{\overline K}}{2p^{2}} - 1} =
ycm^{2} ( 1 + \delta \ln (cm^{2}a^{2}) ),
\end{equation}
where $\delta = (\pi\beta{\overline K}/2p^{2}) - 1$.
The divergent part of the second order term in $y$ has no $q^{2}$-independent
part. When we expand $e^{iq\cdot x}$ in powers of $q$, the odd terms in $q$
vanish when integrated over angles.
To second order in $\delta$ and $y$, the coefficient of $q^{2}$ 
in $\Gamma^{(2)}_{\phi\phi}$ is
\begin{equation}
\frac{(ycm^{2})^{2}}{4}\int d^{2}x  
        x^{2}[\sinh G_{0}(x) - G_{0}(x)]  =  
\frac{(ycm^{2})^{2}}{4}\int_{0}^{m^{-1}}
        \frac{\pi x^{3} dx}{[cm^{2}(x^{2}+a^{2})]^{2+2\delta}}  
+ \mbox{ finite terms}.
\end{equation}
The integral contributes a factor of $\ln a$ :
\begin{equation}
2(cm^{2})^{2}\int_{0}^{m^{-1}}
        \frac{x^{3} dx}{[cm^{2}(x^{2}+a^{2})]^{2+2\delta}}  
= \int_{0}^{m^{-2}}\frac{z dz}{(z+a^{2})^{2}}  
= - \ln (cm^{2}a^{2}).
\end{equation}
The leading divergent contribution to $\Gamma^{(2)}$ is thus
\begin{eqnarray}
\Gamma^{(2)}_{\phi\phi}(q) & = & \frac{p^{2}}{4\pi^{2}\beta{\overline K}}
          (q^{2}+m^{2}) + 
          2ycm^{2}[1+\delta\ln(cm^{2}a^{2})]  \nonumber   \\
   &   & - \frac{\pi}{32}(2y)^{2}q^{2}\ln(cm^{2}a^{2}).
\label{eq:gam-div}
\end{eqnarray}
\par
To establish the RG recursion relations for the hexatic rigidity
$(\beta{\overline K})^{-1}$
and the fugacity $y$, we define renormalized parameters 
$(\beta{\overline K}_{r})^{-1}$ and $y_{r}$ via
\begin{equation}
(\beta{\overline K})^{-1} = Z_{K}(\beta{\overline K}_{r})^{-1} \;\; ,
\;\;\; y = Z_{y}y_{r}.
\label{eq:ren-cons}
\end{equation}
All infinities can be absorbed by two independent renormalization constants.
$Z_{K}$ and $Z_{y}$ can be chosen as functions of $a$ to
ensure that the renormalized two-point vertex,
\begin{equation}
{\Gamma_{R}}^{(2)}_{\phi\phi}(q,y_{r},\delta_{r},\mu) =
\Gamma^{(2)}(q,y,\delta,a),
\end{equation}
is finite for some arbitrary length scale $\mu$, 
order by order in a double expansion in $y_{r}$ and
$\delta_{r} = (\pi\beta{\overline K}_{r}/2p^{2}) - 1$, in the limit
$a \rightarrow 0$.
Thus we expand $Z_{K}$ and $Z_{y}$ in a double series in
$y_{r}$ and $\delta_{r}$. These expansions are substituted into 
Eq. (\ref{eq:ren-cons}), which in turn are inserted in Eq.(\ref{eq:gam-div}).
Then, terms of $\Gamma^{(2)}_{\phi\phi}$ are rearranged by orders 
in the double expansion.
The coefficient in $Z_{K}$ and $Z_{y}$ are determinied by the requirement
that $q^{2}$ and $m^{2}$ have finite coefficients as follows :
\begin{eqnarray}
{\Gamma_{R}}^{(2)}_{\phi\phi} & = & 
  \frac{p^{2}}{4\pi^{2}\beta{\overline K}_{r}Z_{K}^{-1}}(q^{2} + m^{2}) 
  + 2y_{r}Z_{y}cm^{2} 
  \left[1+\left(\frac{\pi\beta{\overline K}_{r}Z_{K}^{-1}}
     {2p^{2}} -1\right)\ln(cm^{2}a^{2}) \right]   
      - \frac{\pi}{32}(2y_{r})^{2}q^{2}\ln(cm^{2}a^{2})
      \nonumber   \\
   & = & \frac{p^{2}}{4\pi^{2}\beta{\overline K}_{r}}(q^{2} + m^{2})
     + 2y_{r}cm^{2} 
   + \left[(Z_{K}-1)\frac{p^{2}}{4\pi^{2}\beta{\overline K}_{r}} -
   \frac{\pi}{2}y_{r}^{2}\ln(cm^{2}a^{2}) \right] q^{2}   \nonumber   \\
   &   & +  2y_{r}\left[(Z_{y}-1) + 
     \left(\frac{\pi\beta{\overline K}_{r}Z_{K}^{-1}}{2p^{2}}
       -1\right)\ln(cm^{2}a^{2})\right]cm^{2} 
\end{eqnarray}
The coefficients $Z_{K}$ and $Z_{y}$ are chosen so that 
${\Gamma_{R}}^{(2)}_{\phi\phi}$ is finite in the limit $a\rightarrow 0$:
\begin{eqnarray}
Z_{K}(\mu) & = & 1 + y_{r}^{2}
     \frac{2\pi^{3}\beta{\overline K}_{r}}
     {p^{2}}\ln \mu^{2}a^{2}  
\label{eq:zyzk}  \\
Z_{y}(\mu) & = & 1 - \left(\frac{\pi\beta{\overline K}_{r}}{2p^{2}}
       -1\right)\ln \mu^{2}a^{2},
\label{eq:zkzy}
\end{eqnarray}
where $\mu$ is an arbitrary mass scale.
From the renormalization constants derived above, we obtain the RG recursion
relations
\begin{eqnarray}
\left. \mu\frac{\partial y_{r}}{\partial\mu}\right|_{{\rm b}} & = & 
   \left. - y_{r}\mu \frac{\partial\ln Z_{y}}{\partial\mu}
   \right|_{{\rm b}} \nonumber  \\
   & = & \left( \frac{\pi\beta{\overline K}_{r}}{p^{2}} - 2 \right) y_{r} ,
\end{eqnarray}
and
\begin{eqnarray}
\left. \mu\frac{\partial (\beta{\overline K}_{r})^{-1}}{\partial\mu}
\right|_{{\rm b}} & = & 
   \left. - (\beta{\overline K}_{r})^{-1}
   \mu \frac{\partial\ln Z_{K}}{\partial\mu}\right|_{{\rm b}} \nonumber  \\
   & = & - \frac{4\pi^{3}}{p^{2}} y_{r}^{2},
\label{krbymu}
\end{eqnarray}
where the subscript {\rm b} means fixed bare parameters.
In terms of length scale $\mu^{-1} = ae^{l}$, these equations are expressed as
\begin{equation}
\frac{d}{dl}(\beta{\overline K})^{-1} =  \frac{4\pi^{3}}{p^{2}} y^{2},  
\end{equation}
\begin{equation}
\frac{d}{dl} y =  \left(2 - \frac{\pi\beta{\overline K}}{p^{2}}\right) y.
\end{equation}
\par
\begin{figure}
\centerline{\psfig{figure=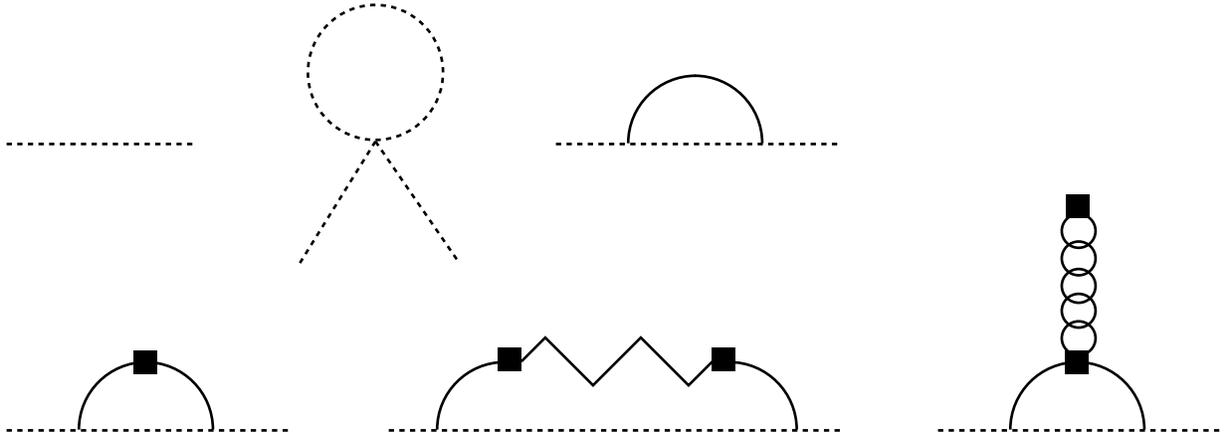}}
\caption{Diagrams contributing to $\Gamma^{(2)}_{hh}$ up to
one loop order and second order in $y$. The dotted line represents the height 
fluctuation field $h$. 
The square dot represents the sum of
tadpole diagrams in Fig. {\protect \ref{fugacity}}, the verticle line of circles
the sum of even numbers of propagators shown in Fig. {\protect \ref{cosh}} and
the sawtooth line the sum of odd numbers of propagators shown in Fig. 
{\protect \ref{sinh}}.
}
\label{hh}
\end{figure}

To complete the RG recursion relations for our Hamiltonian, we study the
renormalization of $\Gamma_{hh}^{(2)}(q)$. We introduce a surface tension term
[Eq. (\ref{sur-ten})] to regularize infrared divergences.
Strictly speaking, we should renormalize $\sigma$ \cite{david-leib,cai-lub}
as well as $\kappa$. However, we will set the renormalized tension
equal to zero in the end. We will, therefore, consider here only
the $q^{4}$ contributions to $\Gamma^{(2)}_{hh}$, which will
determine the renormalization of $\kappa$.
All diagrams contributing to $\Gamma_{hh}^{(2)}$ up to first order in
temperature and second order in $\delta$ and $y$ are shown in Fig. \ref{hh}.
Their sum is written as 
\begin{eqnarray}
\Gamma_{hh}^{(2)}(q) & = & \beta\kappa q^{4} + \frac{3}{8\pi} q^{4}
   \ln(c\sigma a^{2})  
   - \frac{3}{8\pi}\frac{\beta{\overline K}}{4\beta\kappa} q^{4}
   \ln (cm^{2}a^{2})   \nonumber   \\
   &   & - \frac{3}{8\pi}\frac{16\pi^{2}}{p^{2}}
   \frac{(\beta{\overline K})^{2}}{4\beta\kappa} (2y)q^{4}
   (1+\delta\ln(cm^{2}a^{2}))  \nonumber  \\
   &   & - \frac{3}{8\pi}\frac{\pi^{3}}{4p^{2}}
   \frac{(\beta{\overline K})^{2}}{4\beta\kappa} (2y)^{2}q^{4}
   (\ln (cm^{2}a^{2}) )^{2} .
\end{eqnarray}
Defining renormalized parameter $\kappa_{r}$ by
\begin{equation}
\kappa = Z_{\kappa}\kappa_{r},
\end{equation}
and using the renormalization constants $Z_{K}$ and $Z_{y}$ obtained in 
Eq. (\ref{eq:zyzk}) and
Eq. (\ref{eq:zkzy}), we find $\Gamma^{(2)}_{hh}$ is finite in the limit 
$a\rightarrow 0$ if we choose
\begin{eqnarray}
Z_{\kappa}(\mu) & = & 1 - \frac{3}{8\pi}\frac{1}{\beta\kappa_{r}}
   \ln \mu^{2}a^{2}  + 
   \frac{3}{8\pi}\frac{\beta{\overline K}_{r}}{4(\beta\kappa_{r})^{2}} 
   \ln \mu^{2}a^{2}  \nonumber   \\
   &   & - \frac{3}{8\pi}\frac{\pi^{3}}{4p^{2}}
   \frac{(\beta{\overline K}_{r})^{2}}{4(\beta\kappa_{r})^{2}} 
   (2y_{r})^{2} (\ln \mu^{2}a^{2} )^{2} .
\label{eq:zkappa}
\end{eqnarray}
The renormalization group recursion relation for $\kappa_{r}$ is
\begin{eqnarray}
\left. \mu\frac{\partial\beta\kappa_{r}}{\partial\mu}
\right|_{{\rm b}}  & = & 
 \left. - \beta\kappa_{r} \mu\frac{\partial\ln Z_{\kappa}}{\partial\mu}
   \right|_{{\rm b}} 
   \nonumber   \\
    & = & \frac{3}{4\pi} - \frac{3}{4\pi} 
          \frac{\beta{\overline K}_{r}}{4\beta\kappa_{r}},
\end{eqnarray}
when expressed in terms of $l=\ln(\mu a)^{-1}$
\begin{equation}
\frac{d}{dl}\beta\kappa = - \frac{3}{4\pi} 
  \left( 1 - \frac{\beta{\overline K}}{4\beta\kappa} \right).
\label{kappabyl}
\end{equation}
Here we find $y^{2}$ term in Eq. (\ref{eq:zkappa}) is cancelled by
the term proportional to $y^{2}$ from differentiating $\beta{\overline K}_{r}$
by $\mu$ (See Eq. (\ref{krbymu})). Eq. (\ref{kappabyl}) does not have
the term proportional to $y^{2}$.
\par
Thus the complete RG equations
to first order in
temperature and second order in $\delta$ and $y$
for our Hamiltonian are
\begin{equation}
{d \over dl} (\beta{\overline K} )^{-1}  =  {4 \pi ^3 \over p^2} y^2, 
\label{eq:recur1}
\end{equation}
\begin{equation}
{d \over dl} y = \left( 2 - {\pi \beta{\overline K} \over p^2} \right) y,
\label{eq:recur2} 
\end{equation}
\begin{equation}
{d \over dl} \beta \kappa  =  - {3 \over 4 \pi } \left ( 1 - {\beta
{\overline K} \over 4 \beta \kappa} \right ) .
\label{eq:recur3}
\end{equation}
These equations and crikled-to-crumpled phase transition are
analyzed in the preceding paper.
\section{Asymmetry in the disclination fugacities}
On rigid flat membranes, a single positive disclination and a single negative
disclination have the same energy proportional to $\ln(R/a)$ where
$R$ is the linear dimension of the membrane and $a$ is the size of
the core region of the disclination. However, the plus-minus
disclination energy symmetry present in rigid membranes is broken in
deformable membranes \cite{dr-nelson}.
For deformable membranes, the topological charges $q=\pm1/6$ of the
disclinations can partially be cancelled by the curvature charges due
to the Gaussian curvature, when membranes buckle out
to form a cone or a saddle surface.
Membranes with a single disclination undergo a mechanical buckling transition
from a flat configuration to a cone configuration for a positive disclination
or a saddle configuration for a negative disclination.
The Coulombic interaction between disclinations can be screened by Gaussian
curvatures, and the disclination energies are reduced in deformable 
membranes due to buckling into the third dimension. 
Furthermore, the energy of the cone is lower than that of the saddle,
though both energies are proportional to $\ln(R/a)$ \cite{PL}.
The system creates and destroys disclinations to minimize its free energy,
and, in the hexatic phase,
it will always choose configurations in which there is no overall $\ln
(R/a)$ term in the energy.  In flat membranes, this term can be eliminated
by the condition of charge neutrality.  In membranes whose shape can
change, this term can also be eliminated by an average flat configuration
and the condition of charge neutrality.  We believe that this is the most
likely scenario, however, it is not impossible that there are curved
configurations with an overall unbalanced charge that manage to eliminate
the $\ln (R/a)$ energy.  Here we will consider only charge neutral
configurations.  Local buckling may lead to different fugacities $y_+$ and
$y_-$ for positive and negative disclinations.  However, because of the
constraint of charge neutrality, physical observables such as the free
energy will depend only on the product $y_+ y_-$.
\par
We can generalize the our charge neutral
renormalization calculation to take into account
an asymmetry in positive and negative disclination fugacities.
Using the Hamiltonian in Ref. \cite{jp-lub} with different fugacities
for positive and negative disclinations $y_{+}$ and $y_{-}$, respectively,
we modify the sine-Gordon model as follows:
\begin{equation}
{\cal Z} = \int{\cal D}\phi{\cal D}{\bf R} e^{-\beta{\cal H}_{\kappa}
-\beta{\cal H}_{\phi}-i(p/2\pi)\int d^{2}u\sqrt{g}S\phi},
\end{equation}
where
\begin{equation}
\beta{\cal H}_{\kappa} = \frac{1}{2}\beta\kappa\int d^{2}u\sqrt{g} H^{2},
\end{equation}
and
\begin{equation}
\beta{\cal H}_{\phi} = \frac{1}{2\beta K}\left(\frac{p}{2\pi}\right)
\int d^{2}u\sqrt{g}g^{\alpha\beta}\partial_{\alpha}\phi\partial_{\beta}\phi
-\frac{(y_{+}+y_{-})}{a^{2}}\int d^{2}u\sqrt{g}\cos\phi
-i\frac{(y_{+}-y_{-})}{a^{2}}\int d^{2}u\sqrt{g}\sin\phi.
\end{equation}
Following the same renormalization scheme as above, we find the 
recursion relations for the bending rigidity, the hexatic rigidity, and
the fugacities for positive and negative disclinations:
\begin{equation}
\frac{d}{dl}(\beta{\overline K})^{-1} =  \frac{4\pi^{3}}{p^{2}} y_{+}y_{-},
\end{equation}
\begin{equation}
\frac{d}{dl}y_{\pm} = \left( 2-\frac{\pi\beta{\overline K}}{p^{2}} \right)y_{\pm},
\label{y+}
\end{equation}
\begin{equation}
\frac{d}{dl}\beta\kappa = -\frac{3}{4\pi} \left( 
1-\frac{\beta{\overline K}}{4\beta\kappa} \right).
\end{equation}
Defining $y=\sqrt{y_{+}y_{-}}$ and $z=\sqrt{y_{+}/y_{-}}$, we find
\begin{equation}
\frac{d}{dl}(\beta{\overline K})^{-1} =  \frac{4\pi^{3}}{p^{2}} y^{2},
\end{equation}
\begin{equation}
\frac{d}{dl}y = \left( 2-\frac{\pi\beta{\overline K}}{p^{2}} \right)y,
\end{equation}
\begin{equation}
\frac{d}{dl}z = 0.
\end{equation}
Equations for $y$ and ${\overline K}$ are identical to symmetric case.
Equation for $z$ implies the ratio $y_{+}/y_{-}$ does not change under RG.
Thus in the ordered phase, $y_{+}$ and $y_{-}$ scale to zero and the KT
transition occurs at the same temperature as in symmetric case.
\par
For $T>T_{KT}$, both $y_{+}$ and $y_{-}$ grow. If there is an initial
asymmetry between $y_{+}$ and $y_{-}$, it will persist, and positive
disclinations will be favored over negative ones.
This presumably favors positive Gaussian curvature.
A complete discussion of the fluid phase would then require a more complete
treatment of Gaussian curvature.
The flat phase and the Kosterlitz-Thouless transition are, as we have just seen,
not affected by an asymmetry in positive and negative disclination fugacities.
\par
This work was supported in part by NSF grant No. DMR 91-22645 and by
the Penn Laboratory for research in the Structure of Matter under
NSF grant No. 91-20668.

\input psfig
\newpage
\end{document}